\documentstyle[12pt,epsfig,equation]{article}

\begin{document}
\newcommand{\be}{\begin{equation}}
\newcommand{\ben}{\begin{subequations}}
\newcommand{\een}{\end{subequations}}
\newcommand{\beq}{\begin{eqnarray}}
\newcommand{\eeq}{\end{eqnarray}}
\newcommand{\ee}{\end{equation}}
\newcommand{\s}{\\ \vspace*{-4mm}}
\newcommand{\wt}{\widetilde}
\newcommand{\mchi}{\mbox{$m_{\tilde {\chi}_1^0}$}}
\newcommand{\epem}{\mbox{$e^+ e^-$}}
\newcommand{\mchisq}{m_{\tilde {\chi}_1^0}^2}
\newcommand{\lsp}{\mbox{$\tilde {\chi}_1^0$}}
\newcommand{\tanb}{\mbox{$\tan \! \beta$}}
\newcommand{\gchi}{\widetilde{G}\tilde{\chi}^0}
\newcommand{\grav}{\mbox{$\widetilde{G}$}}
\newcommand{\mg}{\mbox{$m_{\widetilde G}$}}
\renewcommand{\thefootnote}{\fnsymbol{footnote}}

\begin{flushright}
APCTP 97--05 \\
PM 97--05 \\
March 1997\\
\end{flushright}

\vspace*{2.5cm}

\begin{center}
{\Large \bf  Higgs Boson Decays into Light Gravitinos} \\
\vspace{10mm}
{\sc Abdelhak Djouadi$^1$ and Manuel Drees$^2$} \\
\vspace{5mm}
{\it $^1$ Laboratoire de Physique Math\'ematique et Th\'eorique, 
UPRES--A 5032,\\ 
Universit\'e de Montpellier II, F--34095 Montpellier Cedex 5, France.}

\vspace*{3mm}

${}^2${\it APCTP, 207--43 Cheongryangri--dong, Tongdaemun--gu, \\
Seoul 130--012, Korea}

\end{center}
\vspace{10mm}

\begin{abstract}
We study decays of the neutral and charged Higgs bosons of the minimal 
supersymmetric standard model into a very light gravitino and a neutralino 
or chargino. Present experimental constraints already imply that the lighter 
scalar Higgs boson can only have a very small branching ratio into this mode. 
However, if the gravitino mass is below $\sim 0.5$ eV, the heavier neutral
and charged Higgs boson decays into a gravitino can be important or even 
dominant. We discuss the resulting signature for associate production 
of the heavy CP--even and the CP--odd states at future $e^+ e^-$ colliders, 
and comment on Higgs boson production at hadron colliders.

\end{abstract}
\clearpage
\setcounter{page}{1}

\pagestyle{plain}

\subsection*{1. Introduction}

Recently models with a very light gravitino \grav\ \cite{9}, $\mg \leq 
10^{-3}$ eV, have attracted some attention \cite{1,2,3}. This interest 
was originally triggered by the resurgence of models \cite{4} where
supersymmetry breaking is transmitted by gauge interactions to the
``visible sector'' containing the usual gauge, Higgs and matter
superfields, although these models tend to predict $\mg \geq 1$
eV. Certain supergravity models can also naturally accommodate a very
light gravitino \cite{4a}. \s

Another motivation comes from CDF's
observation \cite{5} of a single event with an \epem\ pair, two hard
photons, and missing transverse momentum. The probability for such an
event to come from Standard Model (SM) processes is about $10^{-3}$,
but it can quite easily be explained \cite{6,1} in models with a light
gravitino. It should be admitted that the absence of additional events
with two hard photons and missing transverse momentum \cite{7} is
beginning to cast some doubt \cite{8} on this explanation. Nevertheless
models with a very light gravitino remain an interesting field of study in
their own right \cite{9}. \s

In this Letter we explore the impact that a very light gravitino could have 
on the phenomenology of the Higgs sector of the minimal supersymmetric
standard model (MSSM). Higgs bosons do not couple to pairs of gravitinos,
but they do couple to a gravitino and a neutralino or a chargino. The
corresponding branching ratios are computed in Section 2. Since in this
scenario even the lightest neutralino is unstable, $H \rightarrow 
\grav \tilde{\chi}$ decays give rise to final states containing a hard
photon, missing energy, and additional leptons and/or jets if
$\tilde{\chi} \neq \lsp$. Higgs search strategies may have to be modified
to include such final states. One example, the associate production 
of the heavier CP--even and CP--odd neutral Higgs bosons at a future \epem\ 
collider, is discussed in Section 3. Finally, Section 4 contains a brief 
summary and some conclusions.

\subsection*{2. Decay branching ratios}

As first pointed out by Fayet \cite{9}, the couplings of the ``longitudinal''
(spin 1/2) components of the gravitino to ordinary matter are enhanced
by the inverse of the gravitino mass; if \mg\ is sufficiently small, this
can compensate the suppression by the inverse Planck mass $M_P = 2.4 \cdot
10^{18}$ GeV that appears in all gravitational interactions. In fact, a
longitudinal gravitino is \cite{9,10} nothing but the ``Goldstino'' that
signals the spontaneous breaking of global supersymmetry (SUSY), and whose
coupling are inversely proportional to the SUSY breaking scale
$\Lambda^2 \sim \mg M_P$. Since Goldstino couplings contain momenta of
the external particles, partial widths for decays into final states
containing (longitudinal) gravitinos depend very strongly on the mass of the
decaying particle. \s

Specifically, for the case at hand, one has for the partial decay widths
\be \label{e1}
\Gamma (\Phi \rightarrow \tilde{\chi}_i \grav ) = \frac {1} {48 \pi}
\, \kappa_{i\Phi} \, \frac {m^5_{\Phi}} {m^2_{\widetilde G} M^2_P} \, 
\left[ 1 - \left( \frac {m_{\tilde{\chi}_i}} {m_{\Phi}} \right)^2 \, 
\right]^4,
\ee
where $\Phi$ stands for a neutral or charged Higgs boson and $\tilde{\chi}_i$
is one of the four neutralinos or two charginos of the MSSM. The coupling
factors $\kappa_{i\Phi}$ are given by \cite{1}:
\beq 
\kappa_{ih} &=& \left| N_{i3} \sin \! \alpha - N_{i4} \cos \! \alpha
\right|^2  \nonumber\\
\kappa_{iH} &=& \left| N_{i3} \cos \! \alpha + N_{i4} \sin \! \alpha
\right|^2  \nonumber \\
\kappa_{iA} &=& \left| N_{i3} \sin \! \beta + N_{i4} \cos \! \beta
\right|^2  \nonumber \\
\kappa_{iH^\pm} &=& \left| V_{i2} \right|^2 \cos^2 \beta +
\left| U_{i2} \right|^2 \sin^2 \beta. 
\label{e2}
\eeq 
Following the notation of Ref.~\cite{11}, we have introduced the neutralino
mixing matrix $N$, the two chargino mixing matrices $U$ and $V$, the mixing
angle $\alpha$ in the neutral CP--even Higgs sector, and the ratio of Higgs 
vacuum expectation values \tanb. The structure of eqs.~(\ref{e2}) is easily
understood, since gravitinos only couple to members of the same supermultiplet 
(in current basis). Each term is therefore the product of a higgsino component 
of a chargino or neutralino and the component of the corresponding Higgs 
current eigenstate in the relevant Higgs mass eigenstate.\s

It would appear from eq.~(\ref{e1}) than the partial widths for Higgs to
gravitino decays could be made arbitrarily large by making \mg\ very
small, if $m_{\Phi} > m_{\tilde{\chi}_i}$. However, as stated earlier a
very small \mg\ corresponds to a small SUSY breaking scale $\Lambda$.
As argued in Ref.~\cite{1}, present lower bounds on sparticle masses imply that
$\Lambda$ should be several hundred GeV at least, which corresponds to
$\mg >$ (a few) $10^{-5}$ eV. We therefore chose $\mg = 10^{-4}$ eV in most
of numerical examples. This corresponds to $\Lambda = 650$ GeV, which is
probably quite close to its lower bound in realistic models. Note that
$\Lambda^2$ is the vev of the largest F-- or D--term in the model, which
is usually (much) larger than the squared masses of sparticles in the
visible sector. \s

Furthermore, if $\mg \leq 1$ eV, even the lightest 
sparticle in the visible sector will decay inside the detector \cite{9,1}. 
Here we make the usual assumption that the lightest neutralino \lsp\ is 
this next--to--lightest sparticle, the LSP being of course the gravitino.
Associate $\lsp \grav$ production and \lsp\lsp\ pair production at \epem\
colliders then lead to final states with one or two isolated photons and
missing energy--momentum. No significant excess over SM expectations has
yet been observed in these channels. Following Refs.~\cite{2,3} we
interpret this as implying
\ben \label{e3} \beq
\sigma(\epem \rightarrow \grav \lsp) &< 0.1 \ {\rm pb}&  \ \ \ \ 
(\sqrt{s}=91 \ {\rm GeV}, |\cos \theta_\gamma | \leq 0.7 ); 
\label{e3a} \\
\sigma(\epem \rightarrow \lsp \lsp) & < 0.2 \ {\rm pb}& \ \ \ \
(\sqrt{s}=162 \ {\rm GeV}). 
\label{e3b}
\eeq \een 
Expressions for these cross sections can be found in Refs.~\cite{2} and
\cite{12}, respectively. In particular the LEP1 constraint (\ref{e3a}) is 
very restrictive \cite{2}, if $\mchi < M_Z$. \s

Further constraints on the relevant parameter space can be derived from the
negative outcome of chargino searches. Note that in our scenario chargino
pair production always leads to large amounts of visible energy, since 
\lsp\ decays inside the detector. A small $\tilde{\chi}^\pm_1 - \lsp$ mass
difference is thus no longer problematic for these searches. We have
therefore interpreted recent chargino searches at LEP as implying
\be \label{e4}
m_{\tilde{\chi}^\pm_1} > 80 \ {\rm GeV}. 
\ee
In fact, in this scenario existing Tevatron data most likely imply a 
significantly stronger bound on the chargino mass. The authors of 
Ref.~\cite{1} estimate that the present data samples of the CDF and D0
collaborations should contain several additional events with two hard
photons and substantial missing transverse momentum if 
$m_{\tilde{\chi}^\pm_1} \leq 120$ to 130 GeV. No such events have been
observed \cite{7}. Although no ``official'' experimental bound on
$m_{\tilde{\chi}^\pm_1}$ has yet been published for the light gravitino
scenario, it seems very likely that combinations of parameters giving
$m_{\tilde{\chi}^\pm_1} < 120$ GeV are already excluded. We will therefore
indicate the consequences of replacing the lower bound (\ref{e4}) by 
this more restrictive constraint. \s

\noindent
\begin{figure}[htb]
\vspace*{1.2cm}
\centerline{\epsfig{file=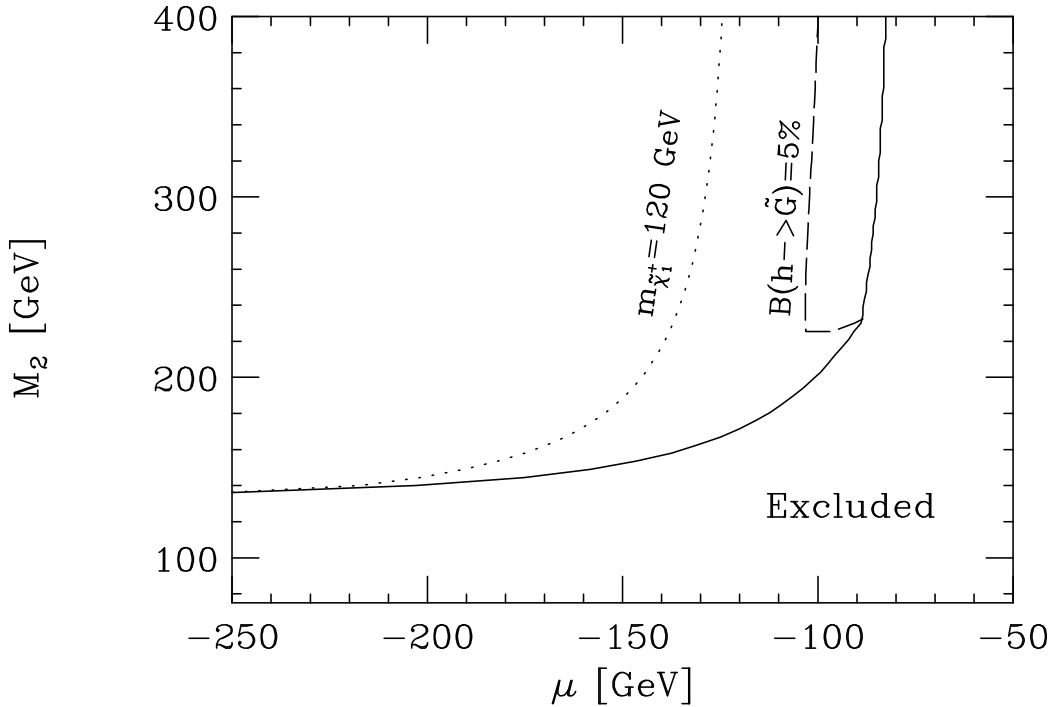,height=10cm}}
\vspace*{-1.7cm}
\caption
{Contour of constant $BR(h \rightarrow \gchi)=5$\% in the
$(\mu,M_2)$ plane, for $m_{\tilde{t}_L}= m_{\tilde{t}_R}=1$ TeV, $A_t =
\sqrt{6}$ TeV, $\tanb=50, \ m_A=500$ GeV and $\mg = 10^{-4}$ eV. The
region to the right and below the solid line is excluded by the experimental
bounds (\ref{e3}) and (\ref{e4}), where we have assumed $m_{\tilde{e}_L} =
m_{\tilde{e}_R}=300$ GeV. In the region to the right and below the dotted
contour the mass of the lighter chargino is below 120 GeV; this region is
most likely excluded by present Tevatron data, as discussed in the text.}
\end{figure}
\vspace*{.7cm}

We are now ready to present some numerical results. We begin with a 
discussion of the decays of the lighter neutral Higgs scalar $h$.
Eq.~(\ref{e1}) shows that $\Gamma(h \rightarrow \gchi)$ will be maximal
if $m_h$ is maximal. In Fig.~1 we have therefore chosen $\tanb = 50$,
a heavy pseudoscalar Higgs boson $m_A = 500$ GeV, large degenerate third
generation squark masses ($m_{\tilde{t}_L} = m_{\tilde{t}_R} = 1$ TeV),
and sizable $\tilde{t}_L - \tilde{t}_R$ mixing ($A_t = \sqrt{6}$ TeV).
This maximizes \cite{13} one--loop corrections to $m_h$, which have been
computed using the formalism of Ref.~\cite{14}; we obtain $m_h = 139$ GeV
for our choice $m_t$(pole) = 176 GeV.\footnote{We include the leading
2--loop QCD corrections \cite{15} by using the running top mass at
scale $Q = \sqrt{m_t m_{\tilde t}}$ in the expressions for the corrected
Higgs boson masses. However, we neglect electroweak corrections \cite{16},
which could reduce $m_h$ by several GeV.} Here and in the subsequent figures
we assume that the $U(1)_Y$ and $SU(2)$ gaugino masses satisfy the 
``unification condition'' $M_1 = 5/3 \tan^2 \theta_W M_2 \simeq 0.5 M_2$.
Relaxing this assumption will not change our results significantly, since
we are interested in the higgsino--like neutralino and chargino states,
see eqs.~(\ref{e2}). Further, we assume $m_{\tilde{e}_L} = m_{\tilde{e}_R} = 
300$ GeV. In this case the constraint (\ref{e3b}) is always weaker than the
combination of the constraints (\ref{e3a}) and (\ref{e4}). Our
calculation of the total decay widths of the MSSM Higgs bosons includes
decays \cite{17} into massive $f \bar{f}$ pairs ($f=c,\  \tau,\ b, \ t$),
loop decays into two gluons or two photons and, where appropriate,
decays into $VV$ final states ($V=W^\pm$ or $Z$, one of which may be 
off--shell) as well as decays of the heavy Higgs bosons into lighter ones
($H \rightarrow hh, \ A \rightarrow Zh$). \s

The region to the right and below the solid line in Fig.~1 is excluded
by present constraints; in the region of small $|\mu|$, the constraint
(\ref{e4}) is the relevant one, while for larger $|\mu|$ (gaugino--like
\lsp) the condition (\ref{e3a}) is most restrictive. This still allows a
narrow strip of parameter space where $BR(h \rightarrow \gchi)$ exceeds
5\%, as indicated by the long--dashed contour. However, this region can
be excluded in the likely case that existing Tevatron data already impose
the limit $m_{\tilde{\chi}^\pm_1} \geq 120$ GeV; this bound is violated to
the right and below the dotted contour. We therefore conclude that, even
if a light gravitino exists, it cannot change the $h$ decay patterns
significantly. \s

\noindent
\begin{figure}[hbt]
\vspace*{1.7cm}
\centerline{\epsfig{file=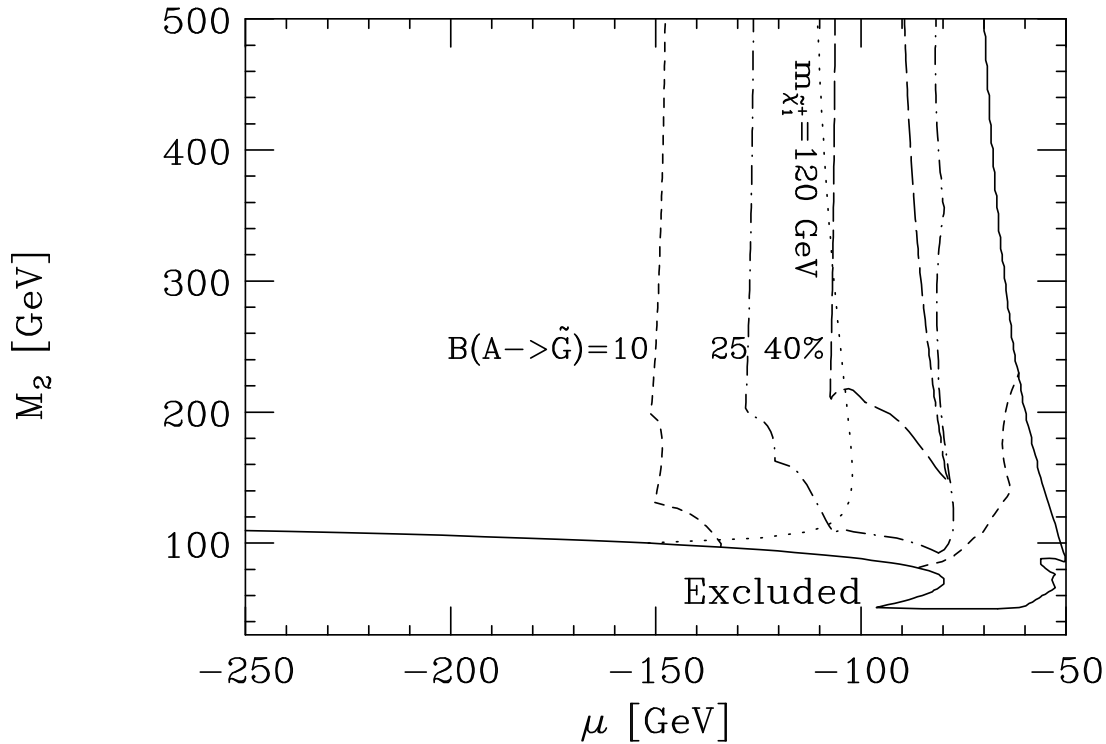,height=10.3cm}}
\vspace*{-1.7cm}
\caption
{Contours of constant $BR(A \rightarrow \gchi)$ in the $(\mu,M_2)$ plane,
for $m_A = 200$ GeV and $\tanb=1.5$. The sfermion and gravitino mass
parameters are as in Fig.~1. The region to the right and below the solid
line is excluded by the constraints (\ref{e3}) and (\ref{e4}), while to
the right and below the dotted contour the lighter chargino mass is below
120 GeV.}
\end{figure}

However, the situation could be quite different for the heavier Higgs
bosons. As an example, in Fig.~2 we show contours of constant $BR( A
\rightarrow \gchi)$ for $\tanb=1.5$ and $m_A=200$ GeV; the values of the
other parameters are as in Fig.~1. We have chosen a small value of \tanb\
here in order to minimize the partial widths for $A \rightarrow b
\bar{b}$ and $A \rightarrow \tau^+ \tau^-$ decays. Since $m_{\tilde t}$ and
$A_t$ are large, $m_h \simeq 100$ GeV so that $A \rightarrow Z h$ decays are
phase space suppressed. We see that in this case the gravitino mode can
account for 40\% of all $A$ decays even if we require $m_{\tilde{\chi}^\pm_1}
\geq 120$ GeV (dotted contour). This is true even though we chose a rather
moderate value of $m_A$ in this figure, so that $AH$ pairs can be produced
at the proposed next linear $\epem$ collider operating at $\sqrt{s}=500$
GeV. \s

Before we discuss $AH$ production in more detail, we explore how the
results of Fig.~2 depend on the values of various parameters. Increasing
\tanb\ tends to reduce the importance of $A$ decays into sparticles, at
least for $m_A < 2 m_t$, since the partial widths into $b \bar{b}$ and
$\tau^+ \tau^-$ pairs scale $\propto \tan^2 \beta$. Choosing the opposite
sign of $\mu$ also reduces the maximal branching ratio for $A \rightarrow
\gchi$ decays. The partial widths for these decays do not change much
when the sign of $\mu$ is flipped, but the constraint (\ref{e4}) is more
restrictive for $\mu > 0$; in addition, the partial widths into pairs of
neutralinos or charginos are enhanced. As a result, we find $BR(A
\rightarrow \gchi) \leq 25$\% for $\mu > 0$, keeping all other parameters
as in Fig.~2; this is reduced to 15\% if we require $m_{\tilde{\chi}^\pm_1}
\geq 120$ GeV. \s

\noindent
\begin{figure}[hbt]
\vspace*{1.4cm}
\centerline{\epsfig{file=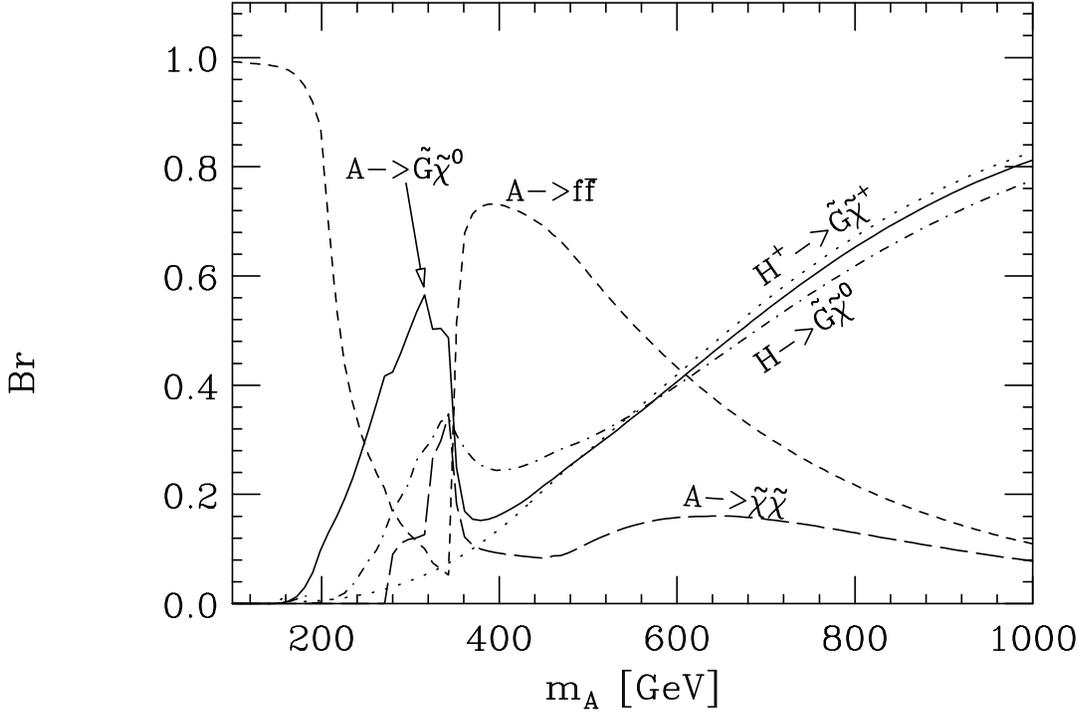,height=10cm}}
\vspace*{-1.7cm}
\caption
{Branching ratios of the heavy Higgs bosons of the MSSM as a 
function of the mass of the pseudoscalar Higgs boson, for $M_2=300$ GeV,
$\mu=-150$ GeV, $\tanb=2, \ \mg=10^{-4}$ eV and
the same sfermion mass parameters as in Fig.~1. Here $f$ stands for
any SM quark or lepton, $\tilde{\chi}^0$ for any of the four neutralinos
of the MSSM, and $\tilde{\chi}$ for any of the two charginos or four
neutralinos.}
\end{figure}

As discussed earlier, we probably cannot enhance the partial widths into
gravitinos by reducing \mg\ even further, since this would give too small
a SUSY breaking scale $\Lambda$. However, we saw in eq.~(\ref{e1}) that
these partial widths also rise very quickly with increasing Higgs boson
mass. This is illustrated in Fig.~3, which shows some branching ratios of
the heavy Higgs bosons of the MSSM as a function of $m_A$. Here we have
chosen $M_2=300$ GeV, $\mu = -150$ GeV and $\tanb=2$, corresponding to
$\mchi=136$ GeV and $m_{\tilde{\chi}^\pm_1}=158$ GeV. The parameters of
the squark mass matrices are as in Figs.~1 and 2. We see that, in spite
of the $m_A^5$ factor, the branching ratio for
$A \rightarrow \gchi$ decays at first grows rather slowly from the
threshold at $m_A=\mchi$, due to the kinematical factor $\left( 1 -
m^2_{\tilde \chi}/m^2_A \right)^4$ in eq.~(\ref{e1}). This is to be
contrasted with the rapid increase of the branching ratio into heavy
SM fermions just beyond the $t \bar t$ threshold (short dashed curve);
here the threshold factor is $\left( 1 - 4 m_t^2 / m_A^2 \right)^{1/2}$.
Finally, the long dashed curve shows the branching ratio for $A$ decays
into pairs of neutralinos or charginos, which exhibits several thresholds.
Together with the onset of $A \rightarrow t \bar{t}$ decays, these
supersymmetric decay modes at first reduce $BR(A \rightarrow \gchi)$ as
$m_A$ is increased beyond 320 GeV. These $\tilde{\chi} \tilde{\chi}$
modes are also responsible for the reduced importance of the gravitino
modes at small $|\mu|$ shown in Fig.~2. However, eventually the $m_A^5$
factor wins out, causing the solid curve to rise again as $m_A$ is
increased beyond 400 GeV. \s

So far we have only discussed decays of $h$ and $A$. For large $m_A$,
$m_A^2 \gg \mu^2$, the partial width for $H \rightarrow \gchi$
decays is very similar to that for $A \rightarrow \gchi$ decays. However,
if \tanb\ is not large, $H$ can also have a sizable branching ratio into
$hh$ pairs \cite{17}. As shown by the dot--dashed curve in Fig.~3, this
reduces the branching ratio for decays of the heavy neutral scalar into
gravitinos. In fact, for the parameters of Fig.~2 this mode completely
dominates $H$ decays, leading to $BR(H \rightarrow \gchi) < 2$\% in this
case. On the other hand, for large $m_A$ the charged Higgs boson has an
even slightly larger branching ratio into the gravitino mode than the 
pseudoscalar (dotted curve). However, at small $m_A$, the branching ratio
for  $H^+ \rightarrow \grav \tilde{\chi}^+$ decays is small, partly because
the lightest chargino is heavier than the lightest neutralino, and partly
because the threshold for $H^+ \rightarrow t \bar{b}$ decays is at much
lower values of $m_A$ than that for $A \rightarrow t \bar{t}$ decays. It
should be emphasized that decays into light gravitinos could {\em dominate}
the decays of all three heavy Higgs bosons of the MSSM, if $m_A \geq 700$
GeV. Note also that for our choice $\mg = 10^{-4}$ eV, the total decay width
of these Higgs bosons is around 100 GeV for $m_A = 1$ TeV. \s

Finally, while it is unlikely that \mg\ is significantly smaller than
$10^{-4}$ eV, it could certainly be larger. We find that for $\mg = 5 
\cdot 10^{-4}$ eV, $BR(A \rightarrow \gchi) < 15$\% for $m_A \leq 1$ TeV.
Clearly this is about the largest value of \mg\ where this branching ratio can 
be sizable, if $m_A$ lies in the range indicated by naturalness arguments.

\subsection*{3. Collider Signals}

The signature for $A \rightarrow \gchi$ decays depends on which neutralino
states are produced in association with the light gravitino. Recall that
these decays proceed through the higgsino components of the neutralinos.
In most cases two of the four neutralinos of the MSSM are higgsino--like
and the other two are gaugino--like, although occasionally two states are
strongly mixed. If the higgsino--dominated states are the heavier
neutralinos $\tilde{\chi}^0_{3,4}$, $A \rightarrow \gchi$ decays will
lead to the subsequent cascade decays $\tilde{\chi}^0_{3,4} \rightarrow
\tilde{\chi}^0_{1,2} f \bar{f}$, where $f$ stands for any kinematically
accessible quark or lepton; these decays proceed through (virtual or
real) $h, \ Z$ and $\tilde{f}$ exchange. $\tilde{\chi}^0_{3,4} 
\rightarrow \tilde{\chi}^\pm_1 f \bar{f}'$ decays may also occur. This can
lead to quite complicated final states. However, in this scenario some
decays of the type $A \rightarrow \tilde{\chi}_{1,2} \tilde{\chi}_{3,4}$
are probably also allowed. Since they combine a higgsino--like and
a gaugino--like state, the relevant couplings are large \cite{11}. In this
case $A \rightarrow \gchi$ decays are therefore only important if $m_A$ is
quite large, probably beyond the reach of next generation colliders. \s

We therefore focus on the scenario where the higgsino--like states are 
lighter than the gaugino--like states. We saw in Fig.~2 that in such a
scenario the branching ratio for $A \rightarrow \gchi$ decays can be
substantial already for $m_A=200$ GeV. Note that the mass splitting between
the higgsino--like states is usually quite small. $A \rightarrow
\grav \tilde{\chi}^0_2$ decays therefore differ from $A \rightarrow
\grav \tilde{\chi}^0_1$ decays only by a rather soft $f \bar{f}$ pair.
With the possible exception $f = \mu$, such soft pairs are probably
not detectable at all at hadron colliders like the planned LHC. At
these colliders the most important Higgs production mechanism is gluon
fusion, $ g g \rightarrow h, \ H, \ A$. The signature for $A \rightarrow
\gchi$ decays would then be a single hard photon from $\lsp \rightarrow
\gamma \grav$ and missing transverse momentum. Note that there is no
Jacobian peak in the $p_T(\gamma)$ distribution, since the photon is
only produced in a secondary decay. Unfortunately, this final state will 
probably suffer from considerable backgrounds\footnote{For instance as 
irreducible backgrounds, one has $\gamma Z$ production 
and subsequent $Z \rightarrow \nu \bar{\nu}$ decay; $W \gamma$ final states 
followed by $W \rightarrow l \nu_l$ ($l=e,\ \mu, \ \tau$) decays can also 
contribute to the background if the lepton is not detected.}. 
Recall also that in this scenario {\em all} SUSY processes give rise to
two hard photons and missing $p_T$ \cite{1,8}. They can therefore also
contribute to the background if one of the photons is lost.\footnote{For
$\mg = 10^{-4}$ eV, $\tilde{q} \rightarrow \grav q$ decays will
dominate even over $\tilde{q} \rightarrow \tilde{g} q$ decays if 
$m_{\tilde q} > 800$ GeV \cite{1}; similarly, gluinos will dominantly
decay into a gluon and a gravitino if $\mg \leq 10^{-3} \ {\rm eV} \cdot
\left( m_{\tilde q} / 1 \ {\rm TeV} \right)^2$. These decays lead to
final states with rather low jet multiplicity, increasing the probability
that the jets escape detection.} A detailed
study will be necessary to decide whether vetoing the presence of
additional jets and/or leptons will be sufficient to overcome this
potentially very large background. \s

As usual, things should be much ``cleaner'' at \epem\ colliders, especially
if $A$ is light enough to be produced at a 500 GeV collider. The main
$A$ production mechanism at \epem\ colliders is associate $AH$ production.
We saw in Sec.~2 that if $m_A \leq 220$ GeV, so that the cross section
for $AH$ production is sizable at a 500 GeV collider, and if \tanb\ is 
relatively small, so that $BR(A \rightarrow \gchi)$ can be substantial, then
$BR(H \rightarrow \gchi)$ is very small, largely due to $H \rightarrow hh$
decays. $AH$ production and subsequent $A \rightarrow \gchi$ decay would
then most often lead to a final state with two $b \bar{b}$ pairs from the
decays of the two light Higgs scalars, a hard isolated photon, and perhaps
a soft $f \bar{f}$ pair from $\tilde{\chi}^0_2 \rightarrow \lsp f \bar{f}$
decays. There should be little background in this channel, especially if one
requires several $b-$tags. Furthermore, events of this type should allow to
determine the masses of {\em all} involved (s)particles by studying
kinematic distributions. Clearly $m_h$ can be obtained from $b \bar{b}$
invariant mass distributions. Combinatorial backgrounds can be reduced
by choosing the pairing that gives the most similar values for the two
$b \bar{b}$ invariant masses. Of course, $m_h$ can also be obtained completely
independently from $\epem \rightarrow Z h$ events in this scenario; the
value of $m_h$ derived from the kind of $AH$ events we are studying here
would therefore be a cross--check on the interpretation of the observed
events. \s

In principle one can get $m_H$ from the invariant mass of the $4b$ system,
and $m_A$ as the mass recoiling against this system. However, the 
$b \rightarrow c \rightarrow s$ cascade decays of the four $b$ (anti)quarks
are likely to produce some neutrinos, which will smear out these invariant
mass distributions. Once $m_h$ is known, one might be able to at least
partly correct for this by using constraint fits. Alternatively, one can
obtain $m_A$ and $m_H$, as well as \mchi, from the photon energy spectrum.
Two examples of this spectrum are shown in Fig.~4, for $m_A = 200$ GeV,
$m_H = 220$ GeV, $\sqrt{s}=500$ GeV and $\mchi = 100$ GeV (solid) and
150 GeV (dashed). Here we have treated $\grav \tilde{\chi}^0_2$ final
states the same as $\grav \lsp$ final states; the additional soft $f
\bar{f}$ pair does not distort the $E_\gamma$ spectrum significantly. \s

\noindent
\begin{figure}[hbt]
\vspace*{1.8cm}
\centerline{\epsfig{file=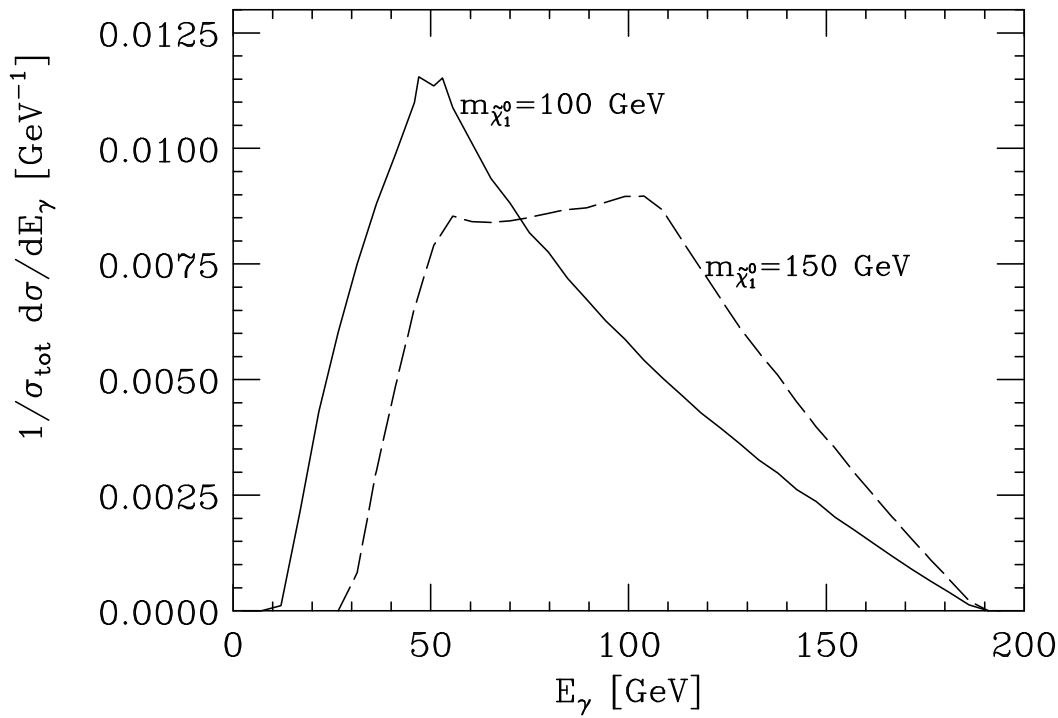,height=10cm}}
\vspace*{-1.7cm}
\caption
{The photon energy spectrum in the lab frame for events of
the type $\epem \rightarrow H A$ with subsequent $A \rightarrow \grav \lsp$
decay and $\lsp \rightarrow \grav \gamma$. We have taken $m_H=220$ GeV,
$m_A=200$ GeV, $\sqrt{s}=500$ GeV, and two different values of the \lsp\ mass, 
as indicated. The acceptance cut $|\cos \! \theta_\gamma | \leq 0.9$ has been
applied.}
\end{figure}
\vspace*{.7cm}

We see that the photons are always quite hard; an acceptance cut like
$E_\gamma > 10$ GeV would therefore not reduce the signal at all. In Fig.~4
we have required $|\cos \! \theta_\gamma| \leq 0.9$, where $\theta_\gamma$
is the angle between the outgoing photon and the $e^-$ beam direction in the
lab frame. This reduces the accepted cross section by some 8\%. Recall
that $d \sigma(\epem \rightarrow AH) / d \cos \! \theta_A \propto 
\sin^2 \theta_A$, where $\theta_A$ is the angle between the pseudoscalar
Higgs boson and the $e^-$ beam; the distribution in $\cos \! \theta_\gamma$
is therefore mildly peaked at $\cos \! \theta_\gamma = 0$. \s

Of course, this angular acceptance cut has no bearing on the endpoints of the
photon energy spectrum, which are at
\be \label{e5}
E_\gamma^{\rm min, max} = \frac{1} {2} \left[ E_{\rm max}(\lsp) \pm
p_{\rm max}(\lsp) \right],
\ee
where $E_{\rm max}(\lsp)$ is the maximal energy of the neutralino in the
lab frame and $p_{\rm max}(\lsp) = \sqrt{ E^2_{\rm max}(\lsp) - 
m^2_{\tilde \chi} }$ is the absolute value of the corresponding 3--momentum.
Since $A \rightarrow \gchi$ decays are isotropic, the distribution in
$E(\lsp)$ is flat, with endpoints
\be \label{e6}
E_{\rm max, min}(\lsp) = \frac {1} {m_A} \left[ E_{(A)}(\lsp) \pm k
E_{(A)}(\grav) \right].
\ee
Here, 
\beq \label{e7} 
E_{(A)}(\lsp) = \frac {m_A^2 + m^2_{\tilde \chi} } { 2 m_A},
\ \ \ \ E_{(A)}(\grav) = \frac {m_A^2 - m^2_{\tilde \chi} } {2 m_A}
\eeq 
are the neutralino and primary gravitino energy in the $A$ rest frame;
$E_A = (s + m_A^2 - m_H^2)/(2 \sqrt{s})$ is the energy of $A$ in the lab 
frame and $k = \sqrt{E_A^2 - m_A^2}$ is the absolute value of the 3--momentum 
of $A$ (or $H$) in the lab frame. \s

The determination of the endpoints of the photon energy spectrum already
gives two constraints on the three unknowns $m_A, \ m_H$ and \mchi. In
addition, Fig.~4 shows that the photon energy spectrum is almost flat over
some range of energies; indeed, it would be completely flat had we not
imposed the acceptance cut $|\cos \! \theta_\gamma | \leq 0.9$. This is due
to the fact that both the $A \rightarrow \gchi$ decay and the $\lsp 
\rightarrow \grav \gamma$ decay are isotropic in their respective rest
frames. In the absence of acceptance cuts  this leads to a flat spectrum
between the points
\be \label{e10}
E_\gamma^\pm = \frac {1}{2} \left[ E_{\rm min} (\lsp) \pm p_{\rm min}
(\lsp) \right],
\ee
where $E_{\rm min}(\lsp)$ has been given in eq.~(\ref{e6}). The solid curve
in Fig.~4 shows that this plateau can be very narrow. This happens for
combinations of parameters leading to a small value of $p_{\rm min} (\lsp)$;
in the given case this quantity amounts to just 6 GeV. In more generic
cases, one should be able to determine both $E^+_\gamma$ and $E^-_\gamma$
with good precision, given sufficient statistics. This would give two
additional, independent constraints on three unknown masses. \s

Another possibly useful kinematic distribution is that in the invariant
mass of the two gravitinos produced in the $A \rightarrow \gchi \rightarrow
\grav \grav \gamma$ decay chain. Since the decays are isotropic, one simply
finds $ d \sigma/ d M_{\widetilde{G} \widetilde{G}} \propto M_{\widetilde{G} 
\widetilde{G}}$ with a sharp cut--off at $M^{\rm max}_{\widetilde{G} 
\widetilde{G}} = \sqrt{m_A^2 - m^2_{\tilde \chi}}$.
However, the $\grav \grav$ invariant mass will be equal to the missing mass
only if there are no additional sources of missing energy in the event
(neutrinos from the decay of $b$ quarks, or initial state radiation that
goes down one of the beam pipes).

\subsection*{4. Summary and Conclusions}

In this note we have studied the decays of MSSM Higgs bosons into light
gravitinos. We found that present experimental bounds already exclude the
possibility that such decays are relevant for the light neutral Higgs
scalar, but they could be quite important, or even dominant, for the heavier 
Higgs bosons if $\mg \leq 0.5$ meV. In the simplest case this gives rise
to final states with a hard isolated photon and missing (transverse) energy.
It is doubtful whether this signal is detectable at the LHC, but associate
$AH$ production at \epem\ colliders will have very little background if
only $A$ decays into a gravitino and a neutralino while $H$ decays into a
fully reconstructable final state. We argued that such a scenario is in 
fact likely if $A$ is sufficiently light to be produced at a 500 GeV collider.
We also showed that in this favorable situation one can in principle
reconstruct $m_H, \ m_A$ and \mchi\ from the energy spectrum of the
photon. \s

In practice it will probably be better to reconstruct the neutralino mass
directly from neutralino pair events. Since the neutralino must have a 
substantial higgsino component if the branching ratio for $A \rightarrow
\gchi$ decays is sizable, the cross section for neutralino pair production at
the same \epem\ collider will be quite substantial -- in fact, significantly
larger than the $AH$ production cross section. This also implies that one
need not worry about Higgs decays into gravitinos if one does not find
sizable signals for neutralino and chargino pair production at the same
energy. In this respect Higgs decays into light gravitinos are quite
similar to Higgs decays into pairs of neutralinos or charginos. At present,
however, all these possibilities must be kept in mind when deriving search
strategies for Higgs bosons at future colliders.

\bigskip

\noindent {\bf Acknowledgements:}
M.D. thanks the members of the LPM at the Universit\'e de Montpellier 
for their hospitality during a visit where this project was initiated.

\end{document}